\documentclass{PoS}

\usepackage{amssymb,amsmath}
\usepackage{amsfonts}
\usepackage{mathrsfs}
\usepackage{graphicx}
\usepackage{subfigure}

\usepackage{amsmath}
\usepackage{amsfonts}
\usepackage{mathrsfs}

\newcommand{\cL}{\mathcal{L}}
\newcommand{\cM}{\mathcal{M}}

\newcommand{\cO}{\mathcal{O}}

\newcommand{\cV}{\mathcal{V}}

\newcommand{\Tr}{{\rm Tr}}

\newcommand{\MeV}{{\rm MeV}}
\newcommand{\GeV}{{\rm GeV}}
\newcommand{\fm}{{\rm fm}}

\newcommand*{\ie}{\textit{i.e.},\ }
\newcommand*{\eg}{\textit{e.g.},\ }

\newcommand*{\et}{\textit{et al.}}

\newcommand{\be}{\begin{equation}}
\newcommand{\ee}{\end{equation}}
\newcommand{\bal}{\begin{align}}
\newcommand{\eal}{\end{align}}

\newcommand*{\chpt}{\raise0.4ex\hbox{$\chi$}PT}
\newcommand*{\schpt}{S\raise0.4ex\hbox{$\chi$}PT}
\newcommand*{\rschpt}{rS\raise0.4ex\hbox{$\chi$}PT}
\newcommand*{\pqchpt}{PQ\raise0.4ex\hbox{$\chi$}PT}
\newcommand*{\pqschpt}{PQ-S\raise0.4ex\hbox{$\chi$}PT}
\newcommand*{\pqrschpt}{PQ-rS\raise0.4ex\hbox{$\chi$}PT}
\newcommand{\SigmaDag}{\Sigma^{\dagger}}
\newcommand{\chiDag}{\chi^{\dagger}}

\newcommand{\dmuSigma}{D_\mu \Sigma}
\newcommand{\dnuSigma}{D_\nu \Sigma}

\newcommand{\dmuSigmaDag}{D_\mu \SigmaDag}
\newcommand{\dnuSigmaDag}{D_\nu \SigmaDag}

\newcommand{\deltapv}{\delta'_V}
\newcommand{\deltapa}{\delta'_A}

\newcommand{\nr}{n_r}
\newcommand{\nrp}{n'_{r}}

\newcommand{\mutwo}{\mu_{(2)}}
\newcommand{\ftwo}{f_{(2)}}

\newcommand{\deltapvtwo}{{\delta'_V}^{(2)}}
\newcommand{\deltapatwo}{{\delta'_A}^{(2)}}

\newcommand{\TildeLpptwo}{\tilde L''_{(2)}}
\newcommand{\TildeLptwo}{\tilde L'_{(2)}}

\newcommand{\denom}{16 \pi^2 \ftwo^2}

\newcommand{\mxvsq}{m_{X_V}^2}

\newcommand{\mxisq}{m_{X_I}^2}
\newcommand{\myisq}{m_{Y_I}^2}
\newcommand{\muisq}{m_{U_I}^2}

\newcommand{\Cc}{\frac{1}{16\pi^2}}

\newcommand{\logmk}{\log\frac{\mu m_s}{\Lambda^2}}
\newcommand{\logmeta}{\log\frac{\frac{4}{3}\mu m_s}{\Lambda^2}}

\def\sl#1{\rlap{\hbox{$\mskip 1 mu /$}}#1}      

\newcommand*{\prd}[1]{Phys.\ Rev.\ {D {\bf #1}}}

\allowdisplaybreaks

\title{Staggered chiral perturbation theory in the two-flavor
case and SU(2) analysis of the MILC data}

\ShortTitle{SU(2) staggered ChPT}

\author{A.~Bazavov$^a$,
C.~Bernard$^b$,
C.~DeTar$^c$,
\speaker{X.~Du$^b$},
W.~Freeman$^a$,
Steven~Gottlieb$^{d,e}$,
U.M.~Heller$^f$,
J.E.~Hetrick$^g$,
J.~Laiho$^h$,
L.~Levkova$^c$,
M.B.~Oktay$^c$,
R.~Sugar$^i$,
D.~Toussaint$^a$,
R.S.~Van~de~Water$^j$ \\ \\
\llap{$^a$}Department of Physics, University of Arizona, Tucson, AZ  85721, USA \\
\llap{$^b$}Department of Physics, Washington University, St.~Louis, MO  63130, USA \\
\llap{$^c$}Physics Department, University of Utah, Salt Lake City, UT  84112, USA \\
\llap{$^d$}Department of Physics, Indiana University, Bloomington, IN  47405, USA \\
\llap{$^e$}National Center for Supercomputing Applications, University of Illinois, Urbana, IL  61801, USA \\
\llap{$^f$}American Physical Society, One Research Road, Ridge, NY  11961, USA \\
\llap{$^g$}Physics Department, University of the Pacific, Stockton, CA  95211, USA \\
\llap{$^h$}SUPA, School of Physics and Astronomy, University of Glasgow, Glasgow G12 8QQ, UK \\
\llap{$^i$}Department of Physics, University of California, Santa Barbara, CA  93106, USA \\
\llap{$^j$}Department of Physics, Brookhaven National Laboratory, Upton, NY  11973, USA}

\abstract{In the light pseudoscalar sector, we study rooted staggered 
chiral perturbation theory in the two-flavor case. The pion mass and decay 
constant are calculated through NLO for a partially-quenched theory. In the 
limit where the strange quark mass is large compared to the light quark masses 
and the taste splittings, we show that the SU(2) staggered chiral theory 
emerges from the SU(3) staggered chiral theory, as expected. Explicit relations between SU(2) and SU(3) low energy constants and taste-violating parameters 
are given. A brief summary of updated SU(2) chiral fits to the MILC lattice 
data is presented.}

\FullConference{The XXVIII International Symposium on Lattice Field Theory, Lattice2010\\
		June 14-19, 2010\\
		Villasimius, Italy}

\begin{document}

\section{Introduction}
\vspace{-0.2cm}
Today most lattice QCD simulations are performed at unphysical light dynamical quark masses. Chiral perturbation 
theory (\chpt)~\cite{GL_SU3,GL_SU2} has proved to be a very important tool for such simulations. By using \chpt, one 
can extrapolate physical quantities to physical light quark masses and get information on low energy 
constants (LECs) of the chiral theory. Although three-flavor \chpt\ has been 
successfully used for simulations with 2+1 dynamical quarks, we are still interested in the two-flavor \chpt\ 
for the following reasons: 1) Usually the simulated light quark masses are much smaller than the simulated strange quark 
mass. We expect the SU(2) expansion to serve as a better approximation and to converge faster than 
the SU(3) one. 2) Fits to SU(2) \chpt\ can give us direction information about LECs in the two-flavor theory. 3)
By comparing results for SU(2) and SU(3) fits, one can study the systematic errors from 
truncations of different versions of \chpt.

In this work, we study the SU(2) \chpt\ for staggered fermions in the partially-quenched case, and 
obtain relations between SU(2) and SU(3) LECs by comparing formulae for the pion mass and decay constant from SU(2) 
and SU(3) \chpt. Then, we perform a systematic NNLO SU(2) chiral analysis for recent MILC data in the light 
pseudoscalar sector. Results for the pion decay constant, SU(2) LECs and chiral condensate in the two-flavor chiral 
limit are presented.
\vspace{-0.2cm}
\section{Rooted SU(2) staggered chiral perturbation theory}
\vspace{-0.2cm}

For lattice simulations based on the staggered fermion formalism, the correct 
effective theory is rooted staggered \chpt\ (\rschpt)~\cite{LEE_SHARPE, AB_M,RUTH_SHARPE_NLO, CB_STAGGERED, BGS_STAGGERED}, 
in which taste-violating effects at finite lattice spacings are incorporated systematically. Physical quantities expressed 
in \rschpt\ become joint expansions in both $m_q$ and $a^2$, where $a$ is the lattice spacing. 
The three-flavor \rschpt\ has been well established~\cite{AB_M} and successfully applied in analyzing the lattice 
data. Here we concentrate on the two-flavor case. Instead of
the three-flavor chiral limit $m_u=m_d=m_s=0$, we perform the expansion around the two-flavor chiral
limit $m_u = m_d = 0, m_s = m_s^{phys}$, where $m_s^{phys}$ is the physical strange quark mass.

The SU(2) \rschpt\ can be constructed by following the same procedure used for SU(3) \rschpt~\cite{AB_M}: 
First, one writes down the Symanzik effective theory (SET) for staggered fermions. Second, 
one maps the terms in the SET to operators in the chiral Lagrangian by using a spurion analysis. The power counting 
rule depends on the specific version of staggered fermions being used. For asqtad staggered fermions, 
we use~\cite{AB_M, MILC_REVIEW} $a^2\delta \sim 2Bm \sim p^2$ where $\delta$ is a typical taste-splitting term. 

At leading order ($\cO(a^2, p^2, m_q)$), the chiral Lagrangian for SU(2) \schpt\ is
    \begin{align}
\cL^{(4)} &= \frac{\ftwo^2}{8} \Tr(\dmuSigma\dmuSigmaDag)
   - \frac{\ftwo^2}{8}\Tr(\chi\SigmaDag + \chi\Sigma) \nonumber \\*
      &+\frac{2m_0^2}{3}(U_I^{11} + \ldots + U_I^{\nr\nr} +D_I^{11} + \ldots +
D_I^{\nr\nr})^2 + a^2\cV, \nonumber \\
\chi &= 2\mutwo Diag( \underbrace{m_x I, \ldots, m_x I}_{\nrp}, \underbrace{m_y
I, \ldots, m_y I}_{\nrp}, \underbrace{m_u I, \ldots, m_u I}_{\nr},
\underbrace{m_d I, \ldots, m_d I}_{\nr}),\label{eq:LOL}
    \end{align}
where $\Sigma = \exp(i\Phi/f)$ and $\cV$ is the LO taste-violating potential. Their definitions can be 
found in Ref.~\cite{DU_SU2}.
In Eq.~(\ref{eq:LOL}), the replica method is used explicitly: we take $n_r'$ copies of each 
valence quark and $n_r$ copies of each sea quark. At the end of the calculations, we set $n_r'=0$ to account 
for partial-quenching, and $n_r=1/4$ for taking the fourth-root of the fermion determinant.



At NLO, the \schpt\ chiral Lagrangian contains two parts: the continuum terms at order 
$\cO(p^4, p^2\\m_q, m_q^2)$ and taste-violating terms at order $\cO(p^2 a^2, m_q a^2, a^4)$. In the 
partially-quenched case, the NLO continuum SU(2) chiral Lagrangian reads:
\begin{align}
    \cL_{cont}^{(6)} &= -\frac{l_1^0}{4}[\Tr (\dmuSigmaDag \dmuSigma)]^2 -
\frac{l_2^0}{4}
    \Tr(\dmuSigmaDag \dnuSigma) \Tr(\dmuSigmaDag \dnuSigma) \nonumber \\*
    &+ p_3^0 \Big( \Tr(\dmuSigmaDag\dmuSigma \dnuSigmaDag\dnuSigma) -
\frac{1}{2}
    [\Tr(\dmuSigmaDag\dmuSigma)]^2 \Big) \nonumber \\*
        &+ p_4^0 \Big( \Tr(\dmuSigmaDag\dnuSigma\dmuSigmaDag\dnuSigma) +
2\Tr(\dmuSigmaDag\dmuSigma\dnuSigmaDag\dnuSigma) \nonumber \\*
        &- \frac{1}{2}[\Tr(\dmuSigmaDag\dmuSigma)]^2 -
\Tr(\dmuSigmaDag\dnuSigma)\Tr(\dmuSigmaDag\dnuSigma) \Big) \nonumber \\*
        &- \frac{l_3^0 + l_4^0}{16}[\Tr(\chi\SigmaDag + \Sigma\chiDag)]^2 +
\frac{l_4^0}{8}\Tr(\dmuSigmaDag\dmuSigma)\Tr(\chi\SigmaDag + \Sigma\chiDag)
\nonumber \\*
        &+ \frac{p_1^0}{16} \Big(\Tr(\dmuSigmaDag\dmuSigma(\chi\SigmaDag +
\Sigma\chiDag)) -
\frac{1}{2}\Tr(\dmuSigmaDag\dmuSigma)\Tr(\chi\SigmaDag + \Sigma\chiDag)
\Big)\nonumber \\*
       &+ \frac{p_2^0}{16} \Big( 2 \Tr(\SigmaDag\chi\SigmaDag\chi +
\Sigma\chiDag\Sigma\chiDag) -
\Tr(\chi\SigmaDag+ \Sigma\chiDag)^2 - \Tr(\chi\SigmaDag - \Sigma\chiDag)^2
\Big) \nonumber \\
       &+ \frac{l_7^0}{16} [\Tr(\chi\SigmaDag - \Sigma\chiDag)]^2\nonumber \\*
       & - l_5^0 \Tr(\SigmaDag{F_R}_{\mu\nu}\Sigma{F_L}_{\mu\nu}) - \frac{i
l_6^0}{2}\Tr({F_L}_{\mu\nu}\dmuSigmaDag\dnuSigma +
{F_R}_{\mu\nu}\dmuSigma\dnuSigmaDag),
\end{align}
where $l_1^0$--$l_7^0$ are bare LECs in ordinary SU(2) \chpt, and $p_1^0$--$p_4^0$ are four new (bare) LECs in 
SU(2) \pqchpt. Operators associated with $p^0_i$ are unphysical operators, in the sense that physical matrix elements of these operators vanish in the limit where the valence quark masses are set equal to sea quark masses.
The specific form of the NLO taste-violating terms are not relevant to this work, so we do not list them here. 
For more details, please see Ref.~\cite{SHARPE_WATER}.  

With the LO and NLO SU(2) \schpt\ Lagrangian, one can calculate the pion mass and pion decay constant 
in the partially-quenched case. Throughout this work, we always assume that the fourth-root procedure is 
legitimate~\cite{BGS_NONLOCAL, MILC_REVIEW}, and in practice it is done by setting $n_r=1/4$ at the end of the calculations. 
The results are
\begin{align}
 \frac{m_{P_5^+}^2}{(m_x + m_y)} = &\mutwo \Big\{ 1 +
\frac{1}{\denom}\Big[\sum_j R_j^{[2,1]} (\{\cM_{XY_I}^{[2]}\}) l(m_j^2)\nonumber
\\*
                &- 2a^2 \deltapvtwo\sum_{j}R_j^{[3,1]}(\{\cM_{XY_V}^{[3]}\})
                         l(m_j^2) + (V\leftrightarrow A) + a^2(\TildeLpptwo +
\TildeLptwo)\Big] \nonumber \\*
                & + \frac{\mutwo}{\ftwo^2}(4l_3+p_1+4p_2)(m_u+m_d) +
                  \frac{\mutwo}{\ftwo^2}(-p_1-4 p_2)(m_x + m_y) \Big\},
\label{eq:mpisq} \\
f_{P_5^+} = &\ftwo \Big\{ 1 + \frac{1}{\denom} \Big[
-\frac{1}{32}\sum_{Q,B}l(m_{Q_B}^2) \nonumber \\*
                       &+ \frac{1}{4} \Big( l(\mxisq) + l(\myisq) +
                    (\muisq-\mxisq)\tilde l(\mxisq) + (\muisq -
\myisq) \tilde l(\myisq) \Big) \nonumber \\*
                       &- \frac{1}{2}\Big(R^{[2,1]}_{X_I}(\{\cM^{[2]}_{XY_I}\})
l(\mxisq)
                        + R^{[2,1]}_{Y_I}(\{\cM^{[2]}_{XY_I}\})l(\myisq)\Big)
\nonumber \\*
                         &+\frac{a^2 \deltapvtwo}{2}\Big(
R^{[2,1]}_{X_V}(\{\cM^{[2]}_{X_V}\}) \tilde l(\mxvsq) + \sum_j D^{[2,1]}_{j,
X_V}(\{\cM^{[2]}_{X_V}\}) l(m^2_j) \nonumber \\*
                                   &+ (X \leftrightarrow Y) + 2\sum_j
R^{[3,1]}_{j}(\{\cM^{[3]}_{XY_V}\})l(m_j^2)\Big) + (V \leftrightarrow
A)\nonumber \\*
                       &+ a^2(\TildeLpptwo - \TildeLptwo) \Big]
+\frac{\mutwo}{2 \ftwo^2}(4l_4 - p_1)(m_u+m_d) + \frac{\mutwo}{2
\ftwo^2}(p_1)(m_x+m_y) \Big\}. \label{eq:fpi}
\end{align}
Here, $\deltapvtwo$ and $\deltapatwo$ are LO taste-violating parameters, and $\TildeLpptwo$ and $\TildeLptwo$ 
are linear combinations of NLO taste-violating parameters. Definitions for meson masses and residue functions 
$R$ and $D$ can be found in Ref.~\cite{DU_SU2}. All LECs in Eqs.(\ref{eq:mpisq}) and (\ref{eq:fpi}) are one-loop renormalized. 

In the limit where the light valence quark masses, light sea quark masses and taste-splittings are all small compared 
to the strange quark mass, \ie $\frac{m_x}{m_s}, \frac{m_y}{m_s}, \frac{m_l}{m_s}, \frac{a^2\Delta_B}{\mu m_s}, 
\frac{a^2\delta'_{V(A)}}{\mu m_s} \sim \epsilon \ll 1$, we expect that SU(2) theory to be generated from the SU(3) 
theory~\cite{GL_SU3}. This can be seen by expanding the corresponding SU(3) formulae~\cite{AB_M} for $m_\pi^2/(m_x+m_y)$ 
and $f_\pi$ in powers of $\epsilon$. Indeed, one can check that the expansion has the same pattern as the SU(2) formulae. 
Furthermore, one can relate SU(3) and SU(2) LECs by comparing these two sets of formulae. First, by comparing LO meson masses 
in various taste channels, we get the relations between LO taste-violating parameters 
    \begin{align}
    a^2\Delta_B^{(2)} = a^2\Delta_B, \ \ \  a^2\deltapvtwo  = a^2 \deltapv, \ \ \ a^2\deltapatwo  = a^2 \deltapa. \label{eq:LOTV}
    \end{align}
Then applying Eq.~(\ref{eq:LOTV}) in the NLO formulae of $m_\pi^2/(m_x + m_y)$ and $f_\pi$, and comparing coefficients of terms 
$m_x, m_y, m_l$ and $a^2$ separately, one obtains the following relations~\cite{DU_SU2}:
\begin{align}
\ftwo &= f (1 - \frac{1}{16\pi^2 f^2} \mu m_s \log\frac{\mu m_s}{\Lambda^2} +
\frac{16L_4}{f^2}\mu m_s), \label{eq:ftwo} \\
\mutwo &= \mu (1 - \frac{1}{48\pi^2 f^2} \frac{4\mu m_s}{3}\log\frac{\frac{4\mu
m_s}{3}}{\Lambda^2} + \frac{32(2L_6-L_4)}{f^2}\mu m_s), \label{eq:mutwo} \\
p_1 &= 16 L_5 - \Cc (1 + \logmk), \label{eq:p1}  \\
p_2 &= -8 L_8 + \Cc
\frac{1}{6}(\logmeta) + \Cc
\frac{1}{4}(1 + \logmk), \label{eq:p2} \\
l_3 &= 8(2L_6 - L_4) + 4(2L_8 - L_5) - \Cc \frac{1}{36}(1 + \logmeta),
\label{eq:l3}  \\
l_4 &= 8L_4 + 4L_5 - \Cc \frac{1}{4} (1 + \logmk), \label{eq:l4} \\
\TildeLpptwo &= \tilde L''
-\frac{1}{6}\Delta_I(1+\logmeta)-\frac{1}{2}\Delta_{av}(1+\logmk), \label{eq:Lpptwo} \\
\TildeLptwo &= \tilde L' -
\frac{1}{6}\Delta_I(1+\logmeta)+\frac{1}{2}\Delta_{av}(1+\logmk),\label{eq:Lptwo}
\end{align}
where $L_4, L_5, L_6$ and $L_8$ are renormalized SU(3) LECs, and $\tilde L''$ and $\tilde L'$
are the NLO taste-violating parameters in SU(3) \rschpt. Equations~(\ref{eq:l3}) and (\ref{eq:l4}) 
are the same as the equations in the full QCD continuum case~\cite{GL_SU3}. Equations~(\ref{eq:p1}) 
and (\ref{eq:p2}) relate the unphysical LECs in the partially-quenched two-flavor theory to 
the physical LECs in the three-flavor theory. Equations~(\ref{eq:Lpptwo}) and (\ref{eq:Lptwo}) give 
us relations between taste-violating parameters in the two-flavor and three-flavor theories. 
\vspace{-0.2cm}
\section{SU(2) chiral analysis of the MILC data}
\vspace{-0.2cm}
Currently, we have gauge ensembles generated from 2+1 dynamical simulations using asqtad staggered fermions. 
Lattice spacings range from $0.15\,\fm$ to $0.045\,\fm$. The light pseudoscalar mass and decay constant are measured 
with different combinations of valence and sea quark masses. With these data, we perform a systematic NNLO 
SU(2) chiral analysis by using the NNLO formulae for $m_\pi^2$ and $f_\pi$. Results of physical quantities are updated 
from the previous analysis in Ref.~\cite{DU_LAT09}.

The NNLO formulae for $m_\pi^2/(m_x+m_y)$ and $f_\pi$ are obtained by combining the NLO formulae Eqs.~(\ref{eq:mpisq}) and 
(\ref{eq:fpi}), possible analytic NNLO terms and the continuum NNLO chiral logarithms provided by Bijnens and L\"ahde~\cite{BIJ_SU2}. 
The root mean square (RMS) average pion mass is used in NNLO chiral logarithms.
In order for the SU(2) formulae to apply, we require both the valence and sea light quark masses to be significantly 
smaller than the strange quark mass. In practice, we used the following cutoff on our data sets: $m_l \le 0.2 m_s,m_x + m_y \le 0.5 m_s,
\mbox{max}(m_x, m_y) < 0.3 m_s$.
Furthermore, in order for the continuum NNLO chiral logarithms to be applicable, we also require the taste-splittings 
between different pion states to be significantly smaller than the kaon and pion masses. The lattices that are at 
least close to satisfying these conditions are fine ($a\approx0.09\,\fm$), superfine ($a\approx 0.06\,\fm$) and ultrafine 
($a\approx 0.045\,\fm$) lattices. In practice, we used superfine and ultrafine lattices for our central value fit, and 
we included fits to all three kinds of lattices to estimate systematic errors.
One difference from the previous fit is that this time we used modified quark masses $m\to \tilde{m} = m + 
a^2/(2\mu)$ in NNLO analytic terms to make the NNLO LECs scale invariant on the lattice, not just in the continuum.
\vspace{-0.2cm}
\section{Results}
\vspace{-0.2cm}
For the central fit, we used three superfine ensembles $(am_l ,am_s) = \{(0.0018,0.018)$, $(0.0025, \\
0.018)$, $(0.0036,0.018)\}$ and one ultrafine ensemble $(am_l ,am_s) = (0.0028,0.014)$. There are a total of 50 data points 
and 30 parameters with appropriate constraints. This fit has a $\chi^2$ of 18 with 20 degrees of freedom, giving 
a confidence level ${\rm CL}\approx 0.6$. The volume dependence at NLO has been included in the fit formulae, and a small 
($\le 0.3\%$) residual finite volume correction~\cite{Bernard:2007ps, MILC_REVIEW} is applied at the end of the calculations.

\begin{figure}
\centering
\subfigure[]{\label{fig:fpi} \includegraphics[width=0.49\textwidth]{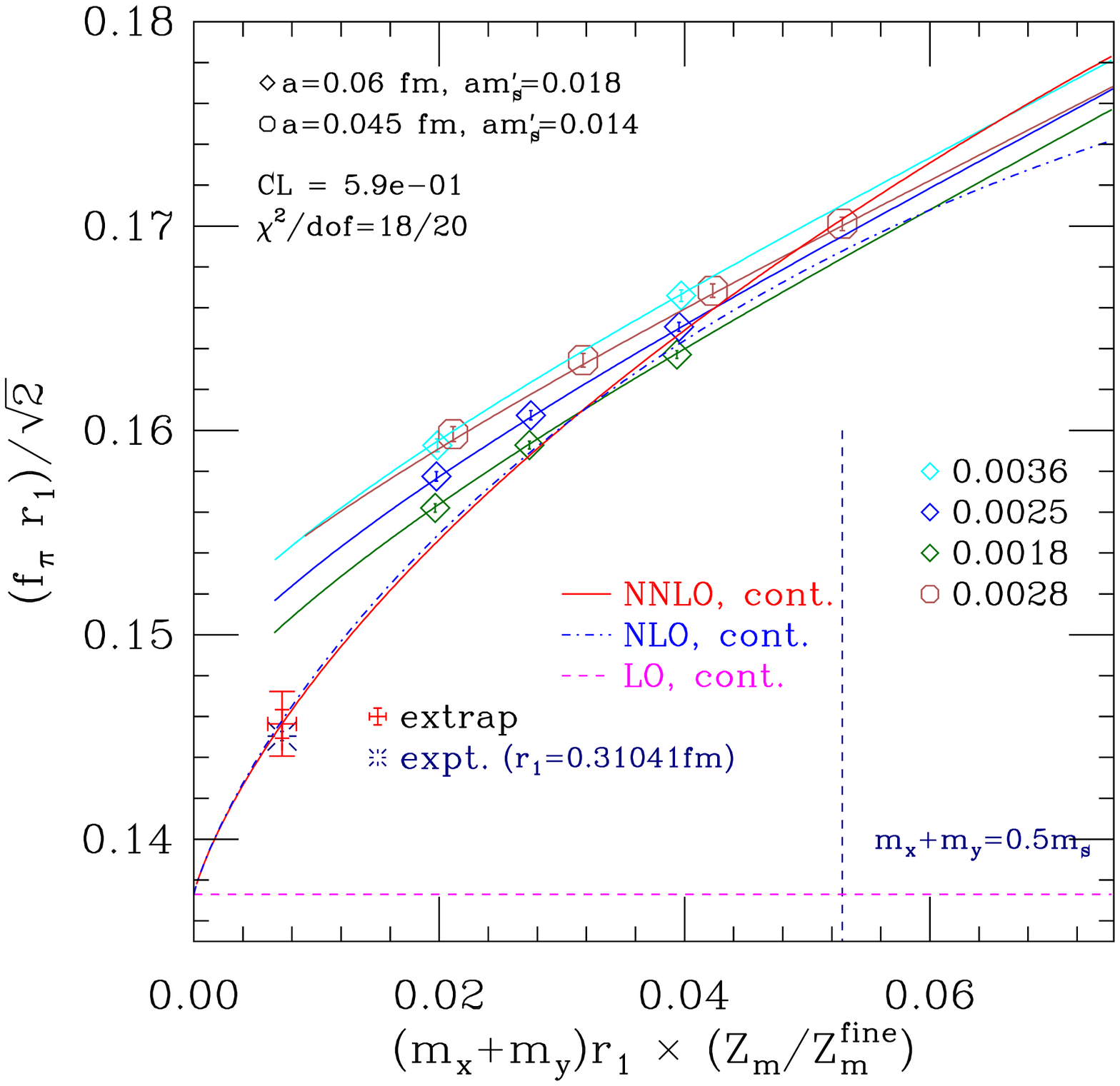}}
\subfigure[]{\label{fig:mpi} \includegraphics[width=0.49\textwidth]{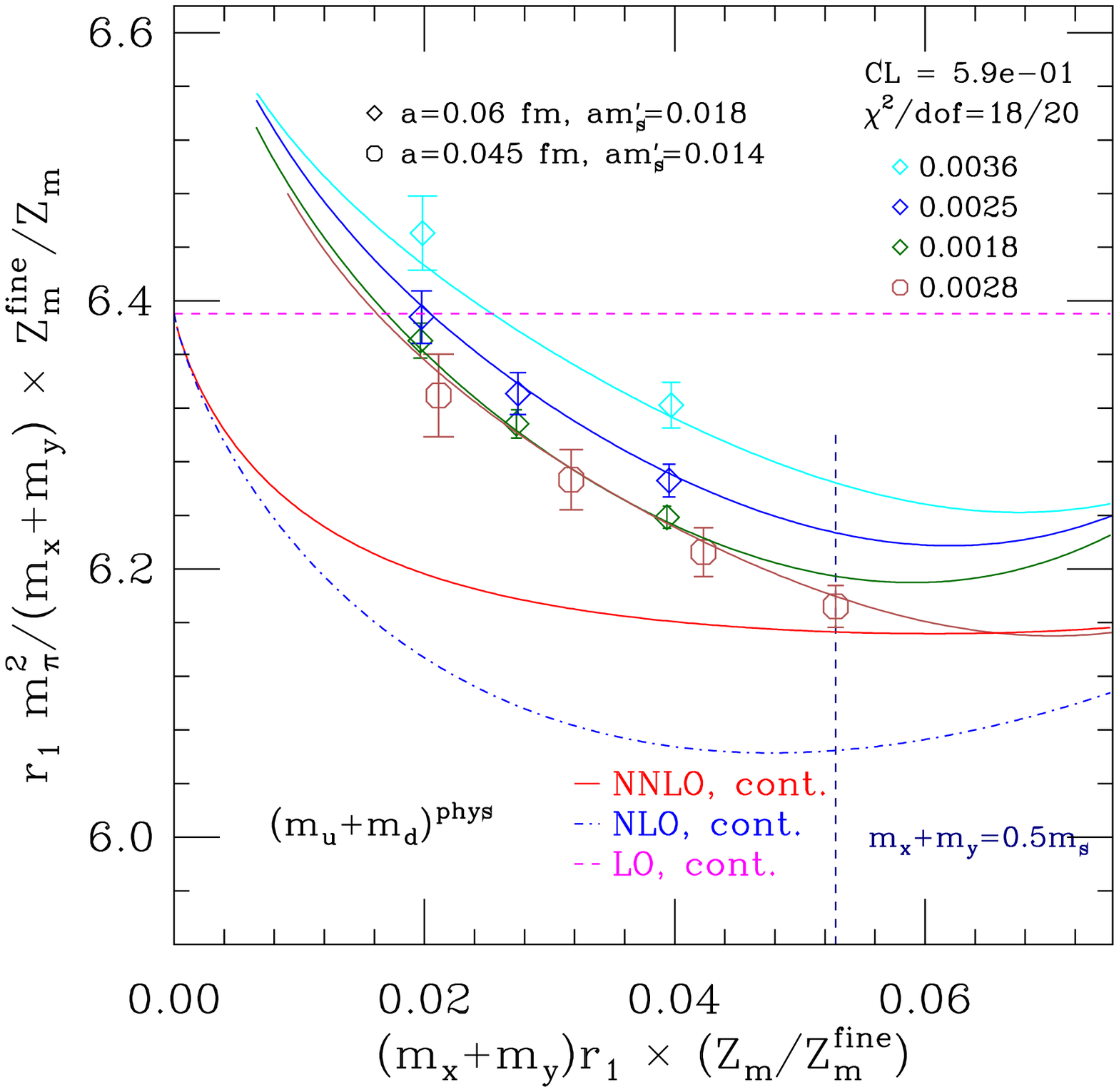}}
\caption{SU(2) chiral fits to $f_\pi$ (left) and $m_\pi^2/(m_x+m_y)$ (right). Only 
points with the valence quark masses equal ($m_x = m_y$) are shown on the plots}
\label{fig:mpisqfpi}
\end{figure}

In Fig.~\ref{fig:mpisqfpi}, we show the fit results for the light pseudoscalar mass and decay constant as functions of 
the sum of the quark masses $(m_x+m_y)$.
The red solid curve represents the full NNLO results for full QCD in the continuum case, where we have set the taste-splittings 
and taste-violating parameters to zero, extrapolated to $a=0$ linearly in $\alpha_s a^2$, and set valence and sea quark masses 
equal. The continuum results through NLO and LO are shown in blue and magenta curves respectively. 

Finally, we find the physical values of the average up and down quark mass $\hat{m}$ by requiring that $\pi$ has its physical 
mass, and then find the decay constant corresponding to this point in Fig.~\ref{fig:fpi}. With the scale $r_1 = 0.3133(23)\,\fm$ 
determined by HPQCD~\cite{HPQCD09}, we obtain $f_\pi = 130.2\pm 1.4 \left(^{+2.0}_{-1.6}\right)\,\MeV$,
where the first error is statistical and the second error is systematic. This agrees with the PDG 2010 value $f_\pi = 130.4\pm0.2\,\MeV$~\cite{PDG}.
Alternatively, one can fix the scale by using the SU(3) NNLO result of $f_\pi$~\cite{CB_LAT10}. We then obtain
    \begin{eqnarray}
    f_2   = 123.8\pm 1.4\left(^{+1.0}_{-3.7}\right)\,\MeV &\qquad&
    B_2   = 2.91(5)(5)(14)\,\MeV  \nonumber\\
    \bar{l}_3  = 2.85\pm 0.81\left(^{+0.37}_{-0.92}\right)  &\qquad&
    \bar{l}_4  = 3.98\pm 0.32\left(^{+0.51}_{-0.28}\right)  \nonumber \\ 
    \hat m =  3.19(4)(5)(16)\,\MeV &\qquad&
    \langle\bar u u\rangle_2  = -[281.5(3.4)\left(^{+2.0}_{-5.9}\right)(4.0)\,\MeV]^3
    \end{eqnarray}
The quark masses and chiral condensate are evaluated in the $\overline{\rm MS}$ scheme at 2 \GeV. We used the two-loop perturbative renormalization 
factor~\cite{Mason:2005bj} to do the conversion. Errors from perturbative calculations are listed as the third errors 
in these quantities. All the quantities agree with SU(3) results~\cite{CB_LAT10} within errors.

\vspace{-0.2cm}
\section{Discussion and outlook}
\vspace{-0.2cm}
In this work, we studied SU(2) \rschpt\ in the partially-quenched case, and we performed a 
systematic SU(2) chiral analysis for recent asqtad data in the light pseudoscalar sector. Results for SU(2) LECs, decay constant and 
chiral condensate in the chiral limit are in good agreement with results from an SU(3) analysis~\cite{CB_LAT10}. It can be seen that the SU(2) 
theory within its applicable region converges much faster than the SU(3) one. For the point $x=0.05$ on the x-axis in Fig.~\ref{fig:mpisqfpi}, 
the ratio of the NNLO correction to the result through NLO is $1\%$ both for $f_\pi$ and $m_\pi^2/(m_x + m_y)$. In contrast, the same ratio 
in SU(3) analysis is $3\%$ for $f_\pi$ and $15\%$ for $m_\pi^2/(m_x + m_y)$, respectively, although the large correction in
the mass case is partly the result of an anomalously small NLO term. 

Since the simulated strange quark masses vary slightly between different ensembles, parameters in SU(2) \chpt\ should also change 
with ensembles. In this work, we also tried to include this effect by using the adjustment formulae in Ref.~\cite{DU_SU2}. It turns out 
that the fits are improved, but not significantly. This part may still need further investigation, hence we do not include these results in 
this work. 
 
In the future, a next step is to include the kaon as a heavy particle in SU(2) \schpt\ in order to study physics involving the strange quark, \eg the kaon mass 
and decay constant. This method has been used in Ref.~\cite{Roessl:1999iu}. Another step is to extend the analysis to data obtained from simulations 
with HISQ fermions, where taste-violating effects are further reduced. This can be done using the same approach as soon as the data are available.

We thank J. Bijnens for providing the FORTRAN code to calculate the NNLO partially-quenched chiral logarithms.

%




\end{document}